\documentclass[aps,twocolumn,prb,floatfix,showpacs,superscriptaddress]{revtex4-1}

\usepackage[utf8]{inputenc}
\usepackage{amsmath,amssymb,amsmath,amsthm,bm,graphicx}
\usepackage{newtxtext,newtxmath} 
\usepackage{booktabs}
\usepackage{placeins}

\usepackage[usenames,dvipsnames]{xcolor} 
\usepackage[colorlinks=true,linkcolor=blue,citecolor=blue,urlcolor=blue]{hyperref} 

\newcommand{\new}[1]{\textcolor{blue}{#1}}

\begin{document}

\author{Michael Seifert}
\email{michael.seifert@uni-jena.de}
\affiliation{Institut f\"ur Festk\"orpertheorie und -optik,
  Friedrich-Schiller-Universit\"at Jena, Max-Wien-Platz 1, 07743 Jena, Germany and European Theoretical Spectroscopy Facility}
 \author{Tom\'{a}\v{s} Rauch }
\email{tomas.rauch@uni-jena.de}
\affiliation{Institut f\"ur Festk\"orpertheorie und -optik,
  Friedrich-Schiller-Universit\"at Jena, Max-Wien-Platz 1, 07743 Jena, Germany and European Theoretical Spectroscopy Facility} 
\author{Miguel A. L. Marques}
\email{miguel.marques@rub.de}
\affiliation{Research Center Future Energy Materials and Systems of the University Alliance Ruhr, Faculty of Mechanical Engineering, Ruhr University Bochum, Universitätsstraße 150, D-44801 Bochum, Germany}
\author{Silvana Botti}
\email{silvana.botti@uni-jena.de}
\affiliation{Institut f\"ur Festk\"orpertheorie und -optik,
  Friedrich-Schiller-Universit\"at Jena, Max-Wien-Platz 1, 07743 Jena, Germany and European Theoretical Spectroscopy Facility}
\affiliation{Research Center Future Energy Materials and Systems, Faculty of Physics and Astronomy, Ruhr Universität Bochum, Germany}

\date{\today}

\title{Structure prediction and characterization of CuI-based ternary $p$-type transparent conductors}

\begin{abstract}
Zincblende copper iodide has attracted significant interest as a potential material for transparent electronics, thanks to its exceptional light transmission capabilities in the visible range and remarkable hole conductivity. However, remaining challenges hinder the utilization of copper iodide's unique properties in real-world applications. To address this, chalcogen doping has emerged as a viable approach to enhance the hole concentration in copper iodide. In search of further strategies to improve and tune the electronic properties of this transparent semiconductor, we investigate the ternary phase diagram of copper and iodine with sulphur or selenium by performing structure prediction calculations using the minima hopping method. As a result, we find 11 structures located on or near the convex hull, 9 of which are unreported. Based on our band structure calculations, it appears that sulphur and selenium are promising candidates for achieving ternary semiconductors suitable as $p$-type transparent conducting materials. Additionally, our study reveals the presence of  unreported phases that exhibit intriguing topological properties. These findings broaden the scope of potential applications for these ternary systems, highlighting the possibility of harnessing their unique electronic characteristics in diverse electronic devices and systems. 
\end{abstract}

\maketitle

\section{Introduction}

Crystalline materials that combine a high electrical conductivity with complete transparency in the visible portion of the electromagnetic spectrum will be crucial components of future transparent electronics and enable innovative technologies, such as transparent electrodes~\cite{Granqvist_1993}, transparent thin-film transistors~\cite{Liu_2018}, solar windows~\cite{Yang_2017}, or electrochromic displays~\cite{Kateb_2016}. The development of transparent displays and transparent electronics, in general, is expected to generate a billion market within the next decade~\cite{AoLiu_2021, Willis_2021}. For such applications both electron and hole transparent conducting materials (TCMs) are needed. However, whereas high-performing $n$-type TCMs, like ZnO \cite{Han_2016} or indium-tin oxide \cite{O'Dwyer_2009, Sakamoto_2018} are already well established, $p$-type counterparts are still lacking \cite{AoLiu_2021, Willis_2021, Hu_2020, AoLiu_2020}.

In this context, CuI has been attracting increasing attention as a multifunctional \textit{p}-type semiconductor \cite{Grundmann_2013}. In fact, the zincblende $\gamma$-phase of CuI displays high hole mobilities ($\mu >40\,\text{cm}^2\text{V}^{-1}\text{s}^{-1}$)~\cite{Yang_2017} and a band gap of around 3.1~eV \cite{Grundmann_2013}. It was furthermore demonstrated that $\gamma$-CuI provides ideal compatibility with numerous $n$-type materials \cite{Yang_2016, Schein_2013, Ding_2012, Lee_2021}. For example, transparent diodes made of $p$-CuI and $n$-ZnO~\cite{Yang_2016} were reported to have rectification ratios as large as $10^9$~($\pm 2\,\text{V}$). In comparison with diodes made of disordered phases, the latter values imply an improvement by two orders of magnitude. Thin-film transistors and displays were successfully established using $p$-type CuI thin films~\cite{Choi_2016,Tixier_2016,Liu_2018,AoLiu_2021}. Possible applications for ultraviolet photodetectors as well as piezoelectric nanogenerators were also discussed \cite{Liu_2016, Yamada_2019,AoLiu_2021}. Measured excitonic binding energies of around 60~meV~\cite{Nikitine_1959} indicate potential for optoelectronic devices: it was indeed proposed to use CuI in blue and UV LEDs~\cite{Ahn_2016, Baek_2020,AoLiu_2021}. Furthermore, CuI thin films were successfully inserted as hole-collecting layers in perovskite solar cells~\cite{Christians_2014,AoLiu_2021}. Promising results regarding photovoltaic efficiency and stability are described in Refs.~\onlinecite{Yu_2018, Matondo_2021}. Thanks to its heavy anion and strong phonon scattering, CuI has also proved potential as transparent thermoelectric material~\cite{AoLiu_2021}. In fact, CuI films are by far the best-known transparent $p$-type thermoelectric with a figure of merit of $ZT=0.21$ at 300~K~\cite{Yang_2017}. Furthermore, in contrast to most used thermoelectric materials, CuI is nontoxic, and therefore suitable for thermoelectric windows, body-heat-driven wearable electronics, as well as on-chip cooling \cite{Yang_2017, AoLiu_2021}. 

Despite of the already good properties of CuI, there is still a need for further improvement before this $p$-type TCM can compete with the $n$-type TCM counterparts. The weak point remains a hole conductivity that is still at least 100 times smaller than the one of $n$-type TCMs. To enhance the conductivity $\sigma$, according to the formula $\sigma = e n_{\rm h} \mu_{\rm h}$  (where $e$ is the electron charge) it is necessary to increase either the hole charge carrier concentration $n_{\rm h}$ or the mobility $\mu_{\rm h}$. In this case, the latter quantities refer to holes (h). The hole mobility, in turn, is proportional to the carrier scattering time and inversely proportional to the hole effective mass. The crystalline quality of the sample plays an important role in increasing the scattering time, while the valence band dispersion yields the transport effective mass. Experimental evidence points to the fact that the film quality can be only marginally improved~\cite{AoLiu_2021}. It is therefore worth considering with priority the route of controlling the hole concentration by doping, as well as reducing the hole effective mass by building CuI-based ternary compounds and alloys. 

Computational screening of suitable dopant and alloy elements is an effective approach for selecting suitable chemical elements for ternary CuI-based materials with the desired optoelectrical properties~\cite{Grauzinyte_2019,Wang_2019,Shi_2017}. Predictive \textit{ab initio} calculations can help experimentalists to prioritize the synthesis and characterization of the most promising materials, uncovering shortcuts to integrate optimized material into new transparent devices. 

It is well known that CuI admits a large range of concentrations of Cu vacancies and that this defect is responsible for $p$-type conductivity in intrinsic samples~\cite{Wang_2011,Huang_2012,Jaschik_2019,Grauzinyte_2019}. Recently, sulfur and selenium substitution of the iodine site were identified as effective shallow acceptors for further $p$-type doping of CuI~\cite{Grauzinyte_2019}. Holes can be easily activated as free charge carriers in the valence band maximum if the impurity energy levels are shallow. Both S and Se substitutions were shown to have low formation energies and lead to higher hole concentrations than Cu vacancies~\cite{Grauzinyte_2019}. We have shown in a recent work that moderate S and Se doping does not modify significantly neither the absorption spectrum of CuI nor hole effective masses and therefore the improved hole concentration can positively impact conductivity without deteriorating transparency~\cite{Seifert_2022_Doping}. Furthermore, doping of CuI with Se~\cite{Storm_2021_Se} and S~\cite{Ahn_2022} was experimentally demonstrated and in the case of S doping also already successfully applied in the context of device fabrication.

It remains an open question if higher concentrations of chalcogen beyond the doping limit, i.e. leading to the formation of a ternary ordered or disordered alloy, can still have beneficial effects on conductivity and transparency. 
In experiments, Se doping of CuI at concentrations larger than 1~at.\% leads to significant changes of the electronic properties~\cite{Storm_2021_Se}, hinting at a phase transition to a ternary material. Ternary phases containing Cu and I are however not included in materials databases~\cite{MP}. To shed light on the possibility to form ordered ternary compounds and disordered alloys, we investigate here the phase diagrams of Cu--S--I and Cu--Se--I, using the minima hopping method (MHM)~\cite{Amsler_2010,Goedecker_2004} with energies and forces determined by density functional theory (DFT). The MHM allows us to search in an unbiased way for new crystal structures, expanding significantly the search space in comparison to previous works on CuI alloys focused on zincblende phases~\cite{Yamada_2020, Seifert_2022_CuBrI, Krueger_2023_AgCuI}. Following this path, we expect to find new materials with promising properties and unveil new ways to control the electronic properties of CuI-based materials. 

Our article is structured as follows. First, we discuss the convex hulls of thermodynamic stability for CuI-based ternaries, uncovering unreported ternary phases. From the inspection of the obtained phase diagrams, we derive a list of candidate stable or quasi-stable ternary systems. These structures are subsequently subjected to further characterization to evaluate their electronic properties using state-of-the-art density functional theory with hybrid functionals. We classify in this way the different structures according to their structural and electronic properties, revealing semiconductors, metals, and topological semimetals. Collecting all this information, we can finally judge the potential of the new compounds for applications as $p$-type transparent conductors. Our analysis discloses the relationship between structural and electronic properties in CuI-based materials, offering precious insight to guide future experimental work.

\begin{figure*}[!htbp]
\begin{center}
\includegraphics[width=\textwidth]{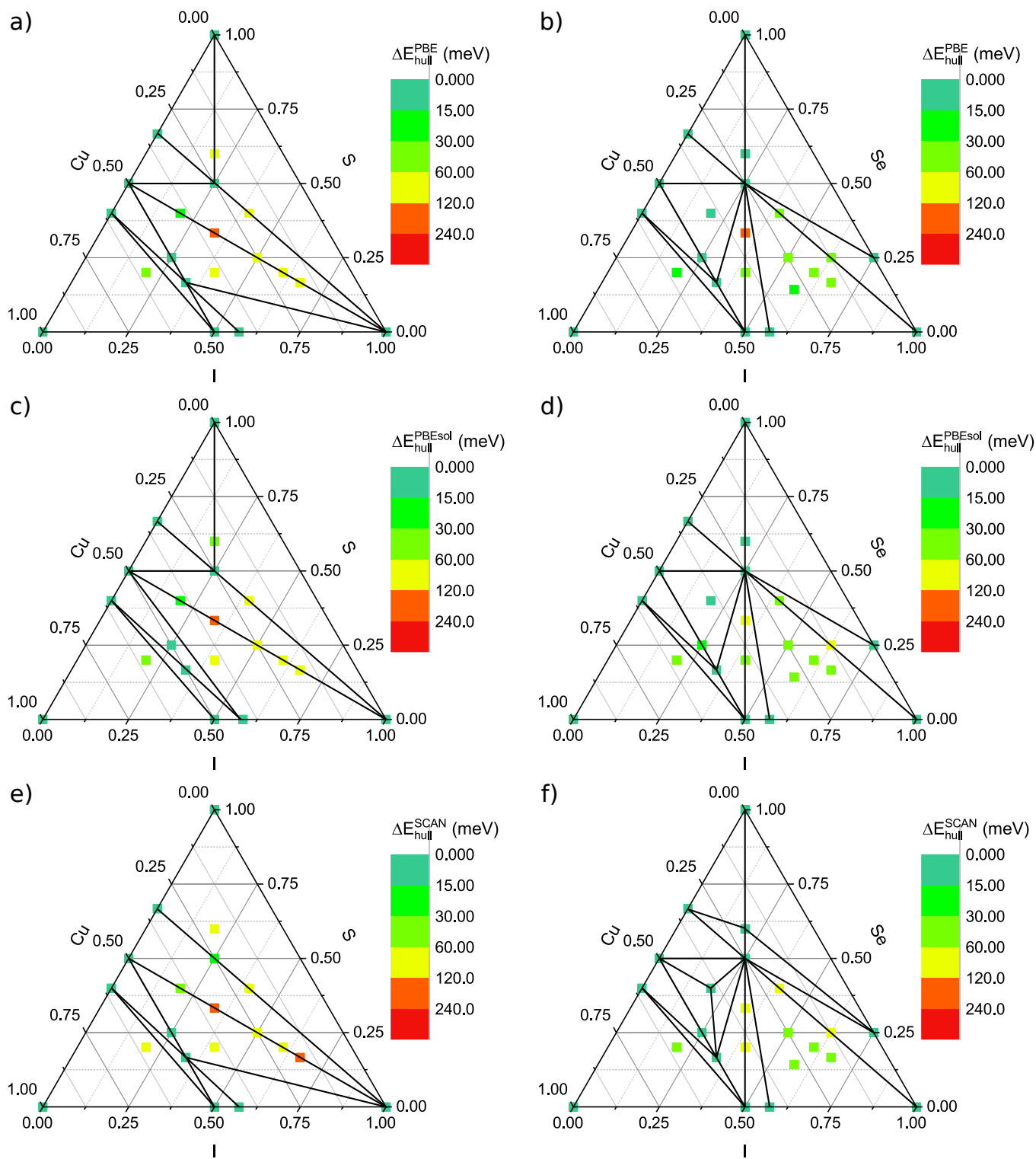} \hfill
\caption{Ternary phase diagrams for Cu--S--I  (panels a, c, e) and Cu--Se--I (panels b, d, f) calculated with DFT and the MHM, as explained in the text. The points denote the compositions that we studied and the colour scheme conveys the energy difference to the convex hull of stability. To evaluate the formation energies on the convex hull different XC-functionals were applied: PBE (a, b), PBEsol (c, d) and SCAN (e, f). Finally, the solid lines connect vertices of the convex hull, i.e., compositions that are thermodynamically stable. As reservoire compounds we used all materials from  Ref.~\onlinecite{Schmidt_2022} and Ref.~\onlinecite{MP}.}
\label{Fig_SCuI_SeCuI_PhaseDiagram}
\end{center}
\end{figure*}

\section{Computational Details}

To calculate the ternary phase diagrams and uncover stable crystalline phases, one has to explore the chemical and configuration space to find the low-lying minima of the adiabatic potential energy surface, considering all possible stochiometries and crystal structures in which atoms can be arranged. At zero temperature, the convex hull of thermodynamic stability is the convex set of low-energy phases that can not decompose thermodynamically into other compounds. To efficiently detect the lowest-energy structures we used the minima hopping method (MHM) \cite{Amsler_2010, Goedecker_2004}. This is a global structure-prediction algorithm designed to determine the most favourable crystal structures of a system given its chemical composition. At a fixed pressure, the enthalpy hypersurface is sampled by performing consecutive short molecular dynamics escape steps followed by local geometry relaxations, taking into account both atomic and cell variables. The initial velocities for the molecular dynamics trajectories are chosen approximately along soft-mode directions, allowing effective escapes from local minima toward low-energy structures. Revisiting already-known crystal structures is avoided by a feedback mechanism. The MHM has already been used for structural prediction in a wide range of materials, including Cu--I binaries~\cite{Jaschik_2019}, with remarkable results~\cite{Huan_2013,Shi_2017,Borlido_2020,SunLin_NatComm_2021,SunLin_2022}.

The MHM relies on an underlying theory to calculate energies and forces. We used to this purpose the framework of DFT, with PAW-pseudopotentials~\cite{Bloechl_1994} (projector augmented wave method) as implemented in the software VASP (Vienna Ab-initio Simulation Package)~\cite{Kresse_1996, kresse_joubert_1999}. The Cu $3p$, $4s$ and $3d$ electrons, I $5s$ and $5p$ electrons, as well as the $s$ and $p$ electrons of the outermost shell of the chalcogen atoms were treated explicitly in the valence. For the structural prediction runs, we used default cutoff values for the plane-wave basis and we choose a ${\bf k}$-point density of 0.025 in reciprocal units. All forces were converged to better than 5~meV/\AA. To approximate the exchange-correlation functional (XC) of density functional theory we used the Perdew-Burke-Ernzerhof~\cite{Perdew_1996} (PBE) generalized gradient approximation.

To avoid biases induced by the starting configurations, the MHM runs were started from random configurations. We performed structural prediction runs for a large selection of stoichiometries, considering unit cells with up to a maximum of 8 atoms. The complete list of considered compositions is given in Table~1 of the Supplementary Information (SI). We remark that in our simulations there is no guarantee that the global minimum is reached, as we cannot exclude that a low-lying minimum is not sampled or that larger unit cells with the same stoichiometry would lead to lower formation energies per atom, e.g. by decreasing their internal energy through geometry distortions. We are moreover neglecting the contribution of entropy that can stabilize disordered alloys rather than ordered crystals at finite temperatures. However, several predictions obtained with this method have been confirmed by experiments (see, e.g., the synthesis of the ferroelectric perovskite LaWN$_3$~\cite{Talley_2021} five years after its prediction~\cite{Sarmiento-Perez_2015}). 

To refine the low-energy crystal structures found during the MHM runs, an additional DFT relaxation step was performed with stricter convergence criteria. To this end, we used a denser \textbf{k}-point grid to satisfy a convergence threshold  of 1~meV/atom and all structures were  relaxed until  forces were smaller than 1~meV/\r{A}. A plane-wave basis set with a cutoff energy of 700~eV was used for all structures, both for this refinement step and for all subsequent electronic-structure calculations.

To build the convex hull, we considered the binary and ternary phases included in the Materials Project database~\cite{MP}, complementing these structures with others that we have identified in previous studies and that are contained in our database~\cite{Schmidt_2022}. We recalculated the energy of all structures with our own set of parameters and convergence criteria for consistency.  For these structures found on the convex hull, or up to 15~meV/atom above it, we repeated the DFT geometry optimization and the calculation of the formation energy using the PBEsol~\cite{Perdew_2008} and SCAN~\cite{SCAN} functionals, with 8000 $k$-points per reciprocal atom. PBEsol is a version of PBE optimized for solids and SCAN is a meta-GGA functional designed to follow 17 exact constraints of the exchange-correlation functional. We considered these functionals in addition to PBE, as it is well known that formations energies can bear significant errors with this widely-used approximation~\cite{Stepanovic_2012, Perez_2015, Tran_2016,Schmidt_2022, Bartel2019}.

To extract the effective masses of the different phases, we considered the symmetry points obtained using SeeK-path~\cite{Hinuma_2017} and we calculated the band structure along these symmetry lines using the PBE approximation. For all semiconducting phases, we further calculated band structures with the hybrid functional HSE06, using an adjusted value of $\alpha$=0.32 to reproduce the experimental band gap of CuI of 3.1~eV~\cite{Seifert_2021}. We refer to the HSE06 with this tuned value of $\alpha$ as ``modified HSE06'' The effective masses were fitted using even polynomials of 6$^{th}$~order, $E(k) = ak^6 + bk^4 + ck^2 + d$, considering an energy range of 25~meV (corresponding to the value of $k_BT$ for T~=~300~K) around the $\Gamma$-point.  
W also calculated the dielectric functions in the independent-particle approximation, i.e. employing the Fermi golden rule and neglecting local field effects. A Lorentzian broadening of 0.1~eV was applied to the spectra. For Cu$_3$SI$_2$ a 11$\times$10$\times$16 \textbf{k}-point grid was used for the calculation of the dielectric function.

\section{Results}
\subsection{Ternary Phase Diagrams}

\begin{table*}[!htbp]
\centering
 \caption{List of crystal structures  that are within 30~meV/atom from the convex hull, corresponding to the points in Fig.~\ref{Fig_SCuI_SeCuI_PhaseDiagram}. For each phase, we indicate composition, space group, formation energy $E_\mathrm{for}^\mathrm{PBE}$ and distance to the convex hull $\Delta E_\mathrm{hull}$ calculated with the functionals PBE, PBEsol and SCAN. For SCAN calculations we also indicate the most favourable decomposition channel if $\Delta E_\mathrm{hull}$ is larger than zero. The space groups were obtained using pymatgen with the ``symprec'' keyword set to 0.1, like it was done in~\cite{Schmidt_2022}.  }
  \label{Table_Ternaries_Energies}
  \begin{tabular}{c | c | c | c | c | c | c }
    \toprule 
           Composition & space group & $E_\mathrm{for}$ (meV/atom) & $\Delta E_\mathrm{hull}^\mathrm{PBE}$ (meV/atom) & $\Delta E_\mathrm{hull}^\mathrm{PBEsol}$ (meV/atom) & $\Delta E_\mathrm{hull}^\mathrm{SCAN}$ (meV/atom) & dissociating to \\
    \midrule \hline
Cu$_3$SI$_2$ & 44 & -179 & 0 & 3 & 4 & Cu$_2$SI + CuI \\
CuS$_2$I & 14 & -118 & 0 & 0 & 23 & - \\ 
Cu$_2$SI & 115 & -180 & 4 & 9 & 11 &  Cu$_3$SI$_2$ + CuS \\
Cu$_2$S$_2$I & 187 & -143 & 18 & 16 & 37 & Cu$_3$SI$_2$ + SCu + CuS$_2$I  \\
CuSe$_2$I & 14 & -133 & 0 & 0 & 0 & - \\
Cu$_3$SeI$_2$ & 8 & -152 & 0 & 4 & 1 & - \\
Cu$_2$SeI & 115 & -137 & 9 & 17 & 11 & Cu$_3$SeI$_2$ + SeCu \\
CuSe$_3$I & 166 & -94 & 12 & 6 & 0 & - \\
Cu$_2$Se$_2$I & 166 & -124 & 12 & 11 & 0 & - \\
Cu$_2$SeI$_4$ & 8 & -85 & 24 & 36 & 38 & CuSe$_2$I + CuI + SeI3 + I  \\
Cu$_3$SeI  & 8 & -100 & 25 & 41 & 35 & CuI + Cu$_3$Se$_2$ + Cu \\
  \end{tabular}
\end{table*}

\begin{table*}[!htbp]
\centering
 \caption{Reaction equations of the ternary compounds that are on the hull using the SCAN functional, in presence of an excess of O$_2$, H$_2$, Cu, I or Se. The reaction energies are calculated using SCAN and are given for the total equation and per formula unit (f.~u.).}
  \label{Table_ReactionEnergies}
  \begin{tabular}{c | c | c | c }
    \toprule  
           Composition &  reaction equation &  total reaction energy (eV) & reaction energy per f.~u. (eV) \\
    \midrule \hline
CuSe$_2$I & 15O$_2$ + 4CuSe$_3$I $\rightarrow$  2Cu(IO$_3$)$_2$ + 2CuSe$_2$O$_5$ + 4SeO$_2$ & -29.114 & -7.279 \\
& 3H$_2$ + CuSe$_2$I $\rightarrow$ HI + 2H$_2$Se + CuH & -4.076 & -4.076 \\
& CuSe$_2$I + Cu $\rightarrow$ Cu$_2$Se$_2$I & -0.215 & -0.215 \\
& CuSe$_2$I + Se $\rightarrow$ CuSe$_3$I & -0.023 & -0.023 \\
& does not decompose with an excess of I & - & - \\
CuSe$_3$I & 19O$_2$ + 4CuSe$_3$I $\rightarrow$  2Cu(IO$_3$)$_2$ + 2CuSe$_2$O$_5$ + 8SeO$_2$ & -38.296 & -9.574 \\
& CuSe$_3$I + H$_2$ $\rightarrow$ H$_2$Se + CuSe$_2$I & -1.611 & -1.611 \\
& 2CuSe$_3$I + Cu $\rightarrow$ CuSe$_2$ + 2CuSe$_2$I & -0.289 & -0.145 \\
& does not decompose with an excess of Se & - & - \\
& CuSe$_3$I + 3I $\rightarrow$ SeI$_3$ + CuSe$_2$I & -0.030 & -0.030 \\
Cu$_2$Se$_2$I & 19O$_2$ + 4Cu$_2$Se$_2$I $\rightarrow$  2Cu(IO$_3$)$_2$ + 2CuSe$_2$O$_5$ + 4CuSeO$_4$ & -36.438 & -9.110 \\ 
 & 7H$_2$ + 2Cu$_2$Se$_2$I $\rightarrow$ 2HI + 4H$_2$Se + 4CuH & -8.874 & -4.437 \\
 & Cu$_2$Se$_2$I + 2Cu $\rightarrow$ Cu$_3$Se$_2$ + CuI & -0.366 & -0.366 \\
 & Cu$_2$Se$_2$I + 3Se $\rightarrow$ CuSe$_2$ + CuSe$_3$I & -0.143 & -0.143 \\
 & Cu$_2$Se$_2$I + I $\rightarrow$ CuI + CuSe$_2$I & -0.048 & -0.048 \\
  \end{tabular}
\end{table*}

\begin{table*}[!htbp]
\centering
 \caption{Structural information from PBE calculations for the crystal structures in Table~\ref{Table_Ternaries_Energies}. We indicate space groups and bond lengths (X stands for S or Se). We also include as a reference the corresponding entries for $\gamma$-CuI. This analysis was carried out using robocrystallographer~\cite{Ganose_2019} and VESTA~\cite{momma_izumi_2011}. }
  \label{Table_Ternaries_StructureAnalysis}
  \begin{tabular}{c | c | c | c | c | c | c }
    \toprule 
           Composition & space group & Cu - I (\r{A}) & Cu - X (\r{A}) & I - X (\r{A}) & X - X (\r{A}) & dimensions  \\
    \midrule \hline
Cu$_3$SI$_2$ & 44 & 2.65 & 2.21 & - & - & 3 \\
 & & 2.74 & 2.25 & - & - & \\
CuS$_2$I & 14 & 2.61 & 2.27 & - &  1.97 & layered \\
Cu$_2$SI & 115 & 2.67 & 2.25 & - & - & 3 \\
Cu$_2$S$_2$I & 187 & 2.62 & 2.29 & - & 2.15 & 3 \\
CuSe$_2$I & 14 & 2.61 & 2.40 & - & 2.31 & layered \\
 & & - & - & - &  2.58 & \\
Cu$_3$SeI$_2$ & 8 & 2.60 & 2.43 & - & - & layered \\
 & & 2.82 & 2.40 & - & - &  \\
 & & 3.01 & - & - & - & \\
Cu$_2$SeI & 115 & 2.66 & 2.37 & - & - & 3 \\
CuSe$_3$I & 166 & 2.68 & 2.42& 3.61 & 2.40 & 3 \\
Cu$_2$Se$_2$I & 166 & 2.59 & 2.41 & - & 2.43 & 3 \\
Cu$_2$SeI$_4$ & 8 & 2.60 & 2.45 & 2.64  & - & layered \\
& & 2.72 & 2.57  &  - & - & \\
Cu$_3$SeI & 8 & 2.60 & 2.61 & - & - & layered \\
 & & 2.81 & 2.53 & - & - & \\
 & & -    & 2.41 & - & - & \\
CuI & 216 & 2.63 & - & - & - & 3 \\
  \end{tabular}
\end{table*}

The ternary phase diagrams calculated with PBE, PBEsol and SCAN are shown in Fig.~\ref{Fig_SCuI_SeCuI_PhaseDiagram} for both Cu--S--I and Cu--Se--I. Additionally to the obtained structures from the MHM run, we added  CuSe$_3$I from the Materials Project database~\cite{Se3CuI_MP, MP}, CuSe$_2$I from Ref.~\onlinecite{Schmidt_2022} and the structure obtained from it  by repacing Se with S, Cu$S_2$I. Those structures could not be obtained in the MHM runs since they have larger unit cells than the ones we considered in our calculations. A full list of the considered stoichiometries is given in Table.~1 of the SI. For each stoichiometry, we found a number of potential structures. In Fig.~\ref{Fig_SCuI_SeCuI_PhaseDiagram} we show the distance to the convex hull of the systems with the lowest energy (i.e. the energetically most favourable phases) at the considered compositions. Regarding binary phases, we show in the phase diagrams just the strictly stable ones extracted from Ref.~\onlinecite{Schmidt_2022} and Ref.~\onlinecite{MP}.

In the case of sulfur, we find a region with several stable phases on the line connecting the stable binaries CuI and CuS. In the case of selenium, this applies as well, but there are additional structures close to the hull. Furthermore, one can see that in the iodine-rich region of the phase diagrams the compounds containing S have a larger distance to the hull than those containing Se. This is already true for the binary phases, and in particular there are no stable phases on the S--I line. 

Comparing the results obtained with the three functionals (PBE, PBEsol and SCAN), we observe only small numerical differences and no qualitative deviations for the convex hulls. Nevertheless, numerical differences can be large enough to change the situation of a compound from being on the hull to being just close to the hull or vice versa. The energy change due to the use of another functional varies from case to case, but performing an arithmetic average for the hull energies of all compounds, the SCAN values are overall the largest and the PBE values the smallest.

The exact values of the formation energies, the energy differences to the convex hull and the most likely dissociation channels are reported in Table~\ref{Table_Ternaries_Energies} for the  ternaries that lie within 30~meV/atom from the convex hull. Concerning the ternary phase diagram including sulfur, we identify two structures, namely Cu$_3$SI$_2$ and CuS$_2$I,  that are located on the convex hull. In the case of the Cu--Se--I phase diagram, beside the corresponding compositions, Cu$_3$SeI$_2$ (but in a distinct structure) and CuSe$_2$I, we find two further stable phases: CuSe$_3$I and Cu$_2$Se$_2$I. The only phase which is exactly on the hull for all functionals is CuSe$_2$I, but as it can be seen in Table~\ref{Table_Ternaries_Energies}, the other phases are either on the hull or very close to it. In total 11 structures appear at a distance of less than 25~meV/atom from the convex hull using the PBE functional. Among them, only 2 were previously reported: CuSe$_3$I~\cite{Se3CuI_MP} and CuSe$_2$I~\cite{Schmidt_2022}.  An overview of the formation energies and distances from the convex hull for all the considered stoichiometries can be also found in Sec.~I the SI. 

For the crystal structures directly on the hull, we calculated reaction energies of the ternary compounds with O$_2$ and H$_2$, as well as with Cu, I and Se using the SCAN functional (see Table~\ref{Table_ReactionEnergies}).  All the structures interact strongly with both oxygen and hydrogen. So it appears extremely important to guarantee an oxygen and hydrogen free environment during the fabrication process. The excess of iodine does not seem to be critical as CuSe$_2$I does not decompose at all and the other two compounds have a small negative reaction energy. On the other hand, in the case of Cu$_2$Se$_2$I the reaction energy in excess of selenium is relatively high. This implies that controlling the right amount of Se for obtaining this specific phase could be a challenge. We remark that these calculations neglect reaction dynamics and are performed at 0~K, neglecting entropic terms.

\subsection{Structural and Electronic Properties}

We now characterize in more detail the structural and electronic properties of the predicted ternary compounds. First, we discuss general trends of the properties of the whole set of Cu--I based materials. Afterwards, in Sec.~\ref{Sec.Discussion}, we subdivide the materials into families with similar structural and electronic properties, to extract from our calculations design rules for  Cu(S,Se)I ternary alloys.

\begin{figure*}[h]
\begin{center}
\includegraphics[width=17cm]{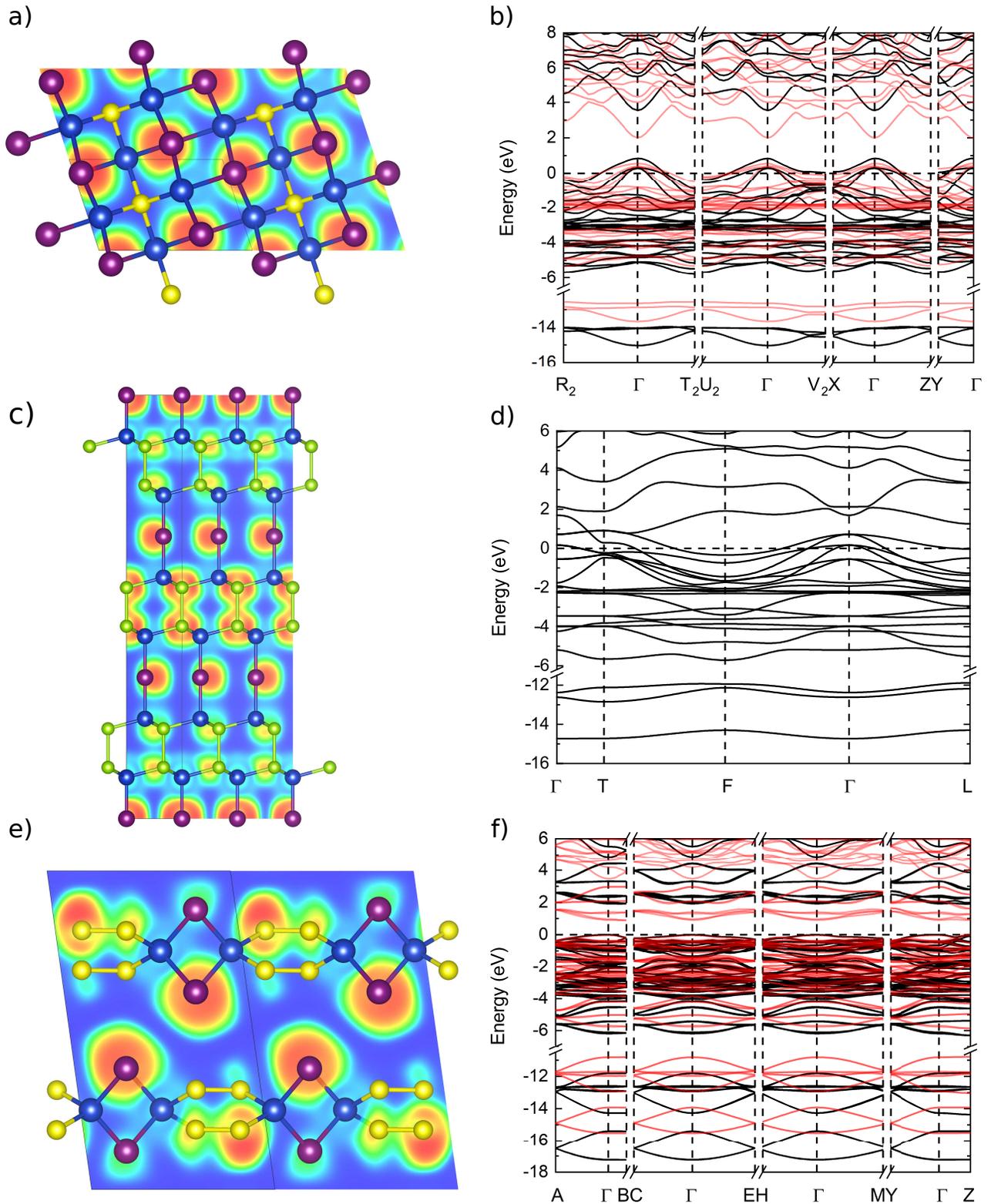} \hfill
\caption{Comparison of the crystal and band structures of the 3 different families of ternary compounds: panel a) and b) Cu$_3$SI$_2$, an example of zincblende-type structures, in panel  c) and d) the topological semimetal Cu$_2$Se$_2$I and in panel e) and f) the Cu-poor indirect semiconductors with chalcogen-chalcogen bonds, exemplified by CuS$_2$I. The figures in panels a), c) e) show the crystal structure and electron-localisation function (ELF) and are produced using VESTA~\cite{momma_izumi_2011}, while panels b), d), f) show the corresponding band structures. The band structures of the semiconductors are calculated using the modified HSE06 functional (black lines) and PBE (red lines), for the semimetal only the PBE band structure is shown.}
\label{Fig_ELFCAR_BS}
\end{center}
\end{figure*}

A summary of the structural parameters is given in Table~\ref{Table_Ternaries_StructureAnalysis}. Pictures of the crystal structures are given in Fig. \ref{Fig_ELFCAR_BS} and in the SI. With exception of Cu$_3$SeI$_2$, all bond lengths between Cu and I atoms are very similar. This holds also for the bonds between Cu and S or Se. The distances between Cu and I are consistently larger than those between Cu and the chalcogen. This can be explained by the larger ionic and covalent radius of I, compared to the corresponding values for Se and S~\cite{Shannon_1976}. Only two structures (CuSe$_3$I and Cu$_2$SeI$_4$) exhibit bonds between a chalcogen and iodine, as  the different anions are usually separated by copper atoms. Another possibility when the number of anions in the formula unit is significantly larger than the number of Cu atoms is the formation of bonds between chalcogen atoms.
Interestingly, both Cu$_3$SeI$_2$ and Cu$_3$SeI crystallize in a layered structure, with Cu atoms in the middle of the layers and anions in the outer part of the layers. These structures are not related to the layered $\beta$-phase of CuI observed by Sakuma~\cite{Sakuma_1988, Seifert_2021}. CuS$_2$I and CuSe$_2$I have also layered structures and present S-S and Se-Se bonds.

\begin{table*}[!htbp]
\centering
 \caption{Calculated PBE and modified HSE06 ($\alpha$=0.32) band gaps $E_\mathrm{g}$ and light hole effective masses in m$_\mathrm{e}$  for the structures in Table~\ref{Table_Ternaries_Energies}. Corresponding values for CuI are given as a comparison. For the non-metallic systems, the contribution of the wavefunctions $\delta$ of the elements at the VBM was calculated using the modified HSE06 ($\alpha$=0.32) to evaluate the p-d-hybridisation.}
  \label{Table_Ternaries_BandGaps}
  \begin{tabular}{c | c | c | c | c | c | c }
    \toprule 
           Composition & PBE (mod. HSE06) $E_\mathrm{g}$ (eV) & PBE $m_\mathrm{eff}^h$ & direction  &  $\delta_\mathrm{Cu}$ & $\delta_\mathrm{I}$ & $\delta_\mathrm{S/Se}$\\
    \midrule \hline
Cu$_3$SI$_2$ & 0 (0) & 0.54 &  $\Gamma$ $\rightarrow$ R$_2$  & 0.488 & 0.301 & 0.210 \\
 & VBM 0.6 (0.8) eV  above $E_\mathrm{F}$  & & & & & \\
 & CBM 2.0 (3.6) eV  above $E_\mathrm{F}$  & & & & & \\
 CuS$_2$I & 1.0 (1.9) (direct) & 0.58 & $\Gamma$ $\rightarrow$ Z & 0.347 & 0.585 & 0.070 \\
 & 0.9 (1.7) (indirect) & & & & & \\
Cu$_2$SI & 0 (0) & 0.47 & $\Gamma$ $\rightarrow$ T$_2$ & 0.475 & 0.230 & 0.293 \\
 & VBM 0.8 (1.3) eV above $E_\mathrm{F}$ & & & & & \\
 & CBM 2.3 (3.9) eV above $E_\mathrm{F}$ & & & & &  \\
 Cu$_2$S$_2$I & 0 & & & & & \\
CuSe$_3$I & 1.0 (2.3) (indirect) & 1.77 & $\Gamma$ $\rightarrow$ A & 0.369 & 0.314 & 0.318 \\
 & 1.2 (2.5) (direct) & & & & & \\
CuSe$_2$I & 0.9 (1.7) (indirect) & 0.78 & $\Gamma$ $\rightarrow$ Z & 0.363 & 0.515 & 0.119 \\ 
 & 0.9 (1.9) (direct) & & & &  \\
 Cu$_3$SeI$_2$ & 0 (0) & 0.52 & $\Gamma$ $\rightarrow$ R$_2$ & 0.459 & 0.184 & 0.356 \\
  & VBM 0.7 (1.1) eV above $E_\mathrm{F}$  & & & & & \\
 & CBM 1.9 (3.3) eV above $E_\mathrm{F}$  & & & & & \\
Cu$_2$Se$_2$I & 0 & & & & & \\
Cu$_2$SeI & 0 (0) & 0.36 & $\Gamma$ $\rightarrow$ T & 0.529 & 0.361 & 0.108 \\
 & VBM 0.7 (0.9) eV above $E_\mathrm{F}$ & & & & & \\
 & CBM 2.0 (3.5) eV above $E_\mathrm{F}$ & & & & & \\
Cu$_2$SeI$_4$ & 0 & & & & & \\
Cu$_3$SeI & 0.4 (1.6) & 0.15 & $\Gamma$ $\rightarrow$ U & 0.504 & 0.023 & 0.472 \\
\hline
CuI & 1.1 (3.1) & 0.21 & $\Gamma$ $\rightarrow$ X & 0.483 & 0.517 & -- \\
  \end{tabular}
\end{table*}

After discussing structural properties, we move to the characterization of electronic properties. To check if the uncovered ternaries might be suitable for applications as \textit{p}-type transparent conductive material we calculate their band structures. A list of band gaps $E_\mathrm{g}$ calculated using the PBE and the modified HSE06 functional (with the parameter $\alpha$=0.32) is shown in Table~\ref{Table_Ternaries_BandGaps}.  We only apply the HSE06 to materials that possess a band gap, specifically Cu$_2$S$_2$I, Cu$_2$Se$_2$I, Cu$_2$SeI$_4$.

Representative band structures are shown in Fig.~\ref{Fig_ELFCAR_BS}, more are shown in Sec.~III of the SI. If we are aiming at applications for transparent electronics, we can observe immediately in panel d) of Fig.~\ref{Fig_ELFCAR_BS} that Cu$_2$SeI$_4$, Cu$_2$Se$_2$I and  Cu$_2$S$_2$I are semimetals and can therefore be filtered out. 

We find then four ternary compounds with 50\% Cu content that are \textit{p}-type degenerate semiconductors: Cu$_3$SI$_2$, Cu$_2$SI, Cu$_3$SeI$_2$ and Cu$_2$SeI. These structures are extremely interesting since they already have intrinsic \textit{p}-type character, as their Fermi energy $E_\mathrm{F}$ is a few hundred meV below the valence band maximum (VBM). We found very similar band structures in stable ordered Cu-I phases with less than 50\% Cu content, namely of Cu$_4$I$_5$ and Cu$_3$I$_4$~ \cite{Jaschik_2019}.  As in the case of Cu-I binaries, the underlying lattice is the zincblende $\gamma$-phase of CuI. However, in that case doping was controlled by Cu vacancies in the binary Cu-I phases~\cite{Jaschik_2019}. In this case, instead, I and chalcogen atoms occupy the  anion site of the underlying (slightly distorted) zincblende lattice alternately. The nature of $p$-type conductivity in these related systems is therefore very different: holes at the top valence of Cu-poor binaries are generated by ordered Cu vacancy complexes, while in Cu-I-based ternaries the ordered $p$-type defects are S or Se substitutions on the I site. We must expect to have in real samples, at finite temperature, disordered configurations of chalcogen atoms on the anion lattice sites, due to entropic effects. The formation of disordered alloys with an underlying zincblende structure is consequently the most likely situation. 

The energy distances between the VBM and conduction band minimum (CBM)  obtained from modified HSE06 calculations are larger than 2.6~eV (only Cu$_3$SeI$_2$ has a smaller band gap of 2.2~eV), yielding a clear indication of  satisfying transparency in the visible spectrum. The suitability as transparent semiconductors largely depends on how deep the Fermi level $E_\mathrm{F}$ lies inside the valence band. In view of applications one has to reach a compromise between the thickness of the ternary layer (that will determine light absorption) and the exact I/chalcogen ratio  (that will determine the concentration of holes and the position of the Fermi energy in the gap).

We can identify finally three indirect semiconductors  with a Cu content lower than 50\% and chalcogen-chalcogen bonds: CuSe$_3$I, CuS$_2$I and CuSe$_2$I. The band structure of CuS$_2$I is shown in panel f) of Fig.~\ref{Fig_ELFCAR_BS} as an example. Note that the PBE value of the direct and indirect band gaps of CuSe$_2$I are the same in Table~\ref{Table_Ternaries_BandGaps}, because the difference is smaller than the number of significant digits. The value of the band gap of these ternaries is always smaller than the gap of CuI. Only CuSe$_3$I with a modified HSE06 direct band gap of 2.5~eV can be expected to be transparent at least for most of the visible spectrum. For the evaluation of  the size of band gaps, we remind that PBE band gaps are usually underestimated by at least 50\%, while hybrid functionals yield reliable predictions \cite{Schilfgaarde_2006, Borlido_2019}. The modification of the mixing parameter of the HSE06 functional was accomplished to match the CuI band gap. As an orientation, the band gap values for CuI are therefore also listed in Table~\ref{Table_Ternaries_BandGaps}. Furthermore, hybrid functionals yield an improved description of localized $d$ states of Cu and therefore of the related $p-d$-hybridisation~\cite{Vidal_2010,Ritter_2020}.  We observe that with the modified HSE functional we obtain values for the projections of the wavefunction at the VBM on atomic orbitals that are in excellent agreement with results from the literature~\cite{Cardona_1963, Seifert_2022_CuBrI} (even if we neglected spin-orbit coupling here). 

We calculated the effective masses of the predicted stable semiconducting materials. Especially important are the effective masses at the top of the valence band, as this quantity is inversely proportional to the mobility of  positive charge carriers. In Table~\ref{Table_Ternaries_BandGaps} we list the smallest light-hole effective masses found with the corresponding direction in the Brillouin zone. The effective masses might depend strongly on the direction, especially for the layered systems. Therefore, and to decide if the material might be suitable as a $p$-type TCM or not, only the best obtained value is displayed here. CuSe$_3$I has an effective mass that is significantly larger than 1, indicating low mobility for positive charge carriers, and can therefore be discarded. The other materials have effective masses below 1 so high mobilities for positive charge carriers can be expected. The best result in this regard is the hole effective mass of Cu$_2$SeI (0.36 electron masses) that is very close to the one of CuI. 

\section{Discussion} \label{Sec.Discussion}
After this general overview of the properties of the ternary phases, we take all materials on the convex hull or at a distance from it of less than 30~meV/atom and divide them in families of systems with similar structural and electronic properties. 
We identified for this purpose the following groups: 1) zincblende-type structures, where S or Se atoms replace a fraction of I atoms on the anion site; 2) topological semimetals similar to the $R-3m$ phase of CuI~\cite{Materialsproject_CuI_R-3m,Le_2018};  3) indirect semiconductors with chalcogen-chalcogen bonds; 4) miscellaneous structures.

\subsection{Zincblende-type structures}\label{Disc.ZB}
We present here more in detail the electronic properties of Cu$_2$SI, Cu$_2$SeI and Cu$_3$SI$_2$. These crystalline systems possess a large band gap and small hole effective masses. These properties are inherited from pristine CuI. We can see in Table~\ref{Table_Ternaries_BandGaps}, that a strong $p-d$-hybridisation between the Cu $3d$ and I $5p$ orbitals characterizes these zincblende phases, as well as the parent $\gamma$-CuI crystal~\cite{Cardona_1963,Grundmann_2013, Seifert_2022_CuBrI}. This property is considered to be important --- together with other factors, like the atomic coordination ---  to obtain a strong band dispersion at the VBM~\cite{Williamson_2017}. We further note that the four-fold coordination of $\gamma$-CuI is preserved in these zincblende-type structures. 

\begin{figure}[h]
\begin{center}
\includegraphics[width=7.5cm]{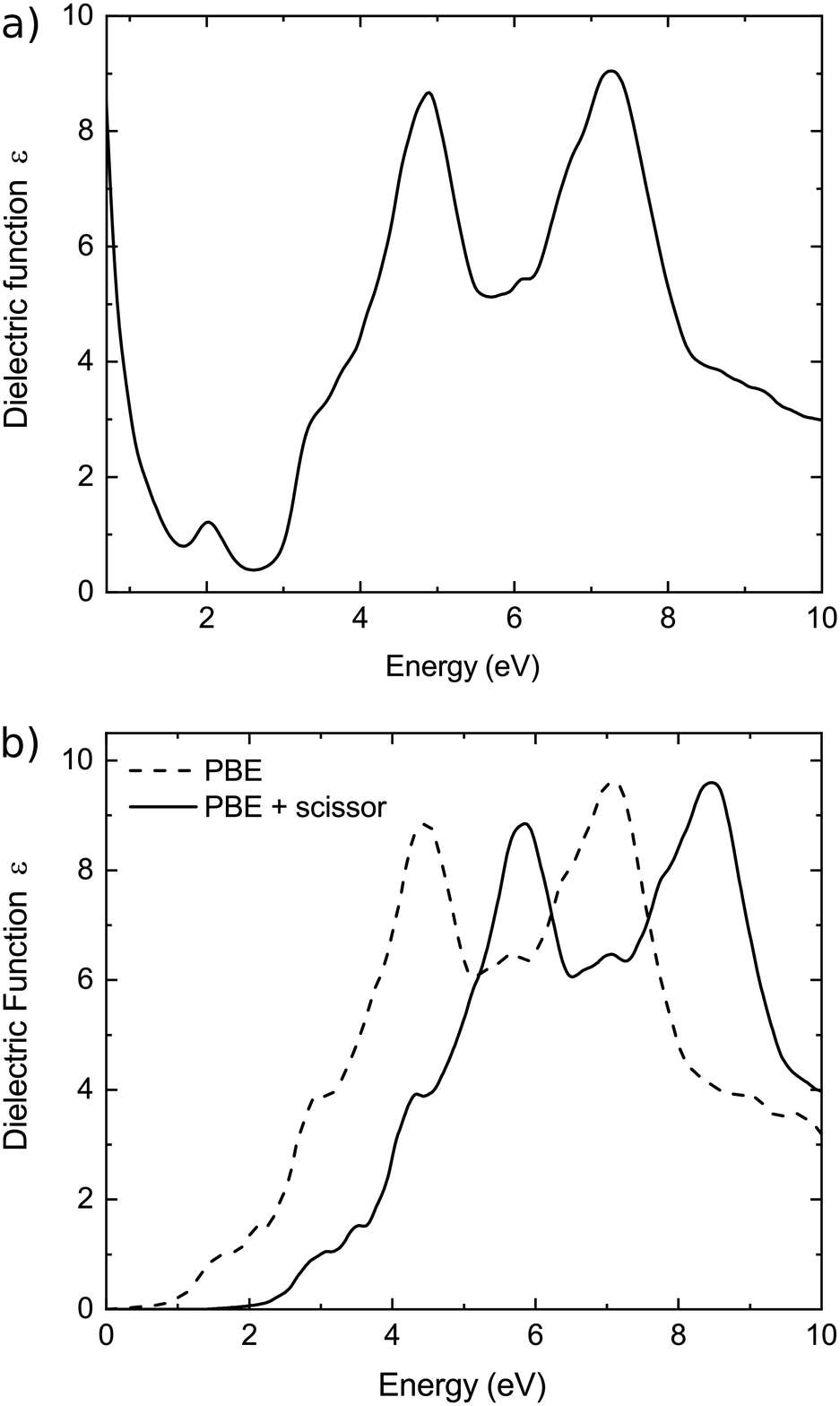} \hfill
\caption{a) Averaged imaginary part of the dielectric function of undoped Cu$_3$SI$_2$, b) Dielectric function of Cu$_3$SI$_2$ calculated with PBE and doped with one electron.  A scissor was applied to PBE to match the modified HSE06 gap. }
\label{Fig_SCu3I2_DielectricFunctionDiagonliased}
\end{center}
\end{figure}

The zincblende-like phase of Cu$_3$SI$_2$ lies on the convex hull and its Fermi energy is close to the VBM (at a distance of \new{0.8~eV}).  We show the electron localization function (ELF) of Cu$_3$SI$_2$ in Fig~\ref{Fig_ELFCAR_BS}. The electrons are, as expected, mainly localized around the anions. A crucial open question concerns the transparency of this phase. The band structures calculated with the PBE and the modified HSE06 functionals are shown in Fig.~\ref{Fig_ELFCAR_BS}. As already discussed, the VBM is above the Fermi level, and therefore we expect light absorption at low frequencies due to valence-valence transitions. For modified-HSE06 calculations, we see that the gap increases significantly as well as the difference between Fermi level and VBM is increased slightly. We calculated the dielectric function of Cu$_3$SI$_2$ and we show it in Fig.~\ref{Fig_SCu3I2_DielectricFunctionDiagonliased}~a).  We can observe at low frequencies, as expected, a Drude-like peak that extends into the visible part of the light spectrum. We conclude that better doping control is necessary to push the Fermi energy closer to the VBM. To illustrate the effect of possible doping we added in Fig~\ref{Fig_SCu3I2_DielectricFunctionDiagonliased}~b) an electron to the unit cell. We see that this cancels the absorption due to valence - valence band transitions completely. Furthermore we added a scissor to take the different gaps from PBE and HSE06 into account. We see that then we observe the desired situation for transparence in the visible. A possible way to reduce the absorption in the gap could be to dope the crystal by substituting donor atoms on the Cu-site. Alternatively, one could tune by alloying the sulfur concentration (and therefore the carrier density of the material) through the whole composition range. S-doping of CuI was proved to be possible and CuI, CuS and Cu$_3$SI$_2$ are all zincblende-type structures on the convex hull. Furthermore, also Cu$_2$SI is very close to the hull. So the fabrication of a Cu$_{x+y}$S$_x$I$_y$ alloy should be possible over a wide range by just tuning the concentrations $x$ and $y$. This means that the hole densities can be tuned as desired. As we argued in the introduction, and according to the discussion in Ref.~\cite{Raj_2019} concerning TiO$_2$ incorporation in CuI, an increased hole density leads to reduced resistivity, which can be beneficial for the application of CuI in electronics.

\subsection{Topological semimetals}

\begin{figure}[h]
\begin{center}
\includegraphics[width=\columnwidth]{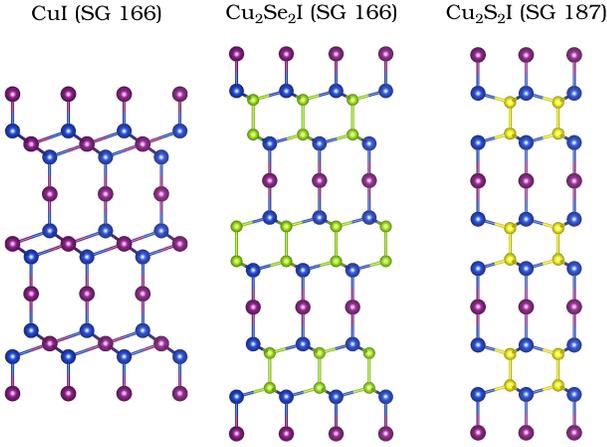} \hfill
\caption{Atomic structures of CuI (SG 166)~\cite{Materialsproject_CuI_R-3m,Le_2018}, Cu$_2$Se$_2$I (SG 166), and Cu$_2$S$_2$I (SG 187).}
\label{FigTop_structures}
\end{center}
\end{figure}

We continue by discussing the unusual electronic properties of the structures belonging to the second family: Cu$_2$Se$_2$I and Cu$_2$S$_2$I.  These crystals are both semimetals and do not possess a band gap close to the Fermi level. The atomic structures of the two compounds resemble closely that of the $R-3m$ phase of CuI~\cite{Materialsproject_CuI_R-3m,Le_2018}. We present all three structures in Fig.~\ref{FigTop_structures} for comparison. Although Cu$_2$Se$_2$I and Cu$_2$S$_2$I look very similar, they do not belong to the same space group (SG). Cu$_2$Se$_2$I has the trigonal SG 166 (like the $R-3m$ phase of CuI) and Cu$_2$S$_2$I has the hexagonal SG 187. One of the key differences to discuss the details of the electronic band structures of the two compounds is the presence of inversion symmetry in Cu$_2$Se$_2$I (SG 166) and its absence in Cu$_2$S$_2$I (SG 187).

\begin{figure}[h]
\begin{center}
\includegraphics[width=\columnwidth]{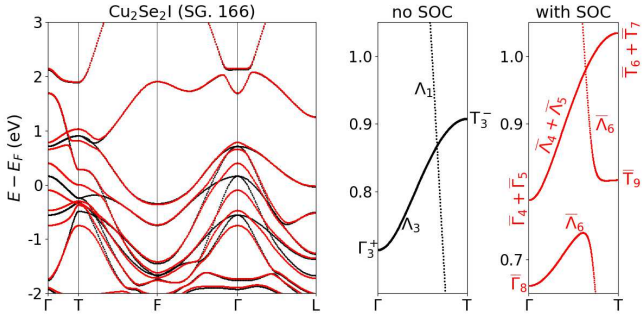}
\caption{Electronic band structure of Cu$_2$Se$_2$I without SOC (black) and with SOC (red). The bands are calculated using the PBE XC functional.}
\label{FigCu2Se2I_bands}
\end{center}
\end{figure}

The $R-3m$ phase of CuI has previously been identified as a Dirac semimetal with Dirac points located on the $\Gamma$-T line at the Fermi level~\cite{Le_2018}$^,$~\footnote{In Ref.~\onlinecite{Le_2018} the point T is denoted as Z.}. Cu$_2$Se$_2$I has the same SG as CuI and is a metal: we expect hence that is possibly exhibits similar features in its electronic band structure shown in Fig.~\ref{FigCu2Se2I_bands}. Indeed, without the inclusion of spin-orbit coupling (SOC) we find a band crossing along the $\Gamma$-T line $\sim$1.0~eV above the Fermi level. We used the IrRep code~\cite{IRAOLA2022108226} and the Bilbao Crystallographic Server~\cite{Elcoro:ks5574} to identify the irreducible representation (IRREPs) of the bands of interest along the $\Gamma$-T line in the Brillouin zone (BZ). As it can be seen in the central panel of Fig.~\ref{FigCu2Se2I_bands}, without considering SOC, the twofold degenerate $\Lambda_3$ band crosses the nondegenerate $\Lambda_1$ band. An avoided crossing of the two bands is not possible due to their different IRREPs along $\Gamma$-T. After the inclusion of SOC, the $\Lambda_3$ band splits into the twofold degenerate $\overline{\Lambda}_6$ and $\overline{\Lambda}_4+\overline{\Lambda}_5$ bands, whereas the $\Lambda_1$ IRREP becomes $\overline{\Lambda}_6$ (twofold degenerate). The crossing of the $\overline{\Lambda}_6$ and $\overline{\Lambda}_4+\overline{\Lambda}_5$ bands is thus protected by the symmetry of the $\Gamma$-T line. Both bands are doubly degenerate (due to the presence of inversion symmetry in the SG 166) and the crossing is thus a Dirac point. 

\begin{figure}[h]
\begin{center}
\includegraphics[width=\columnwidth]{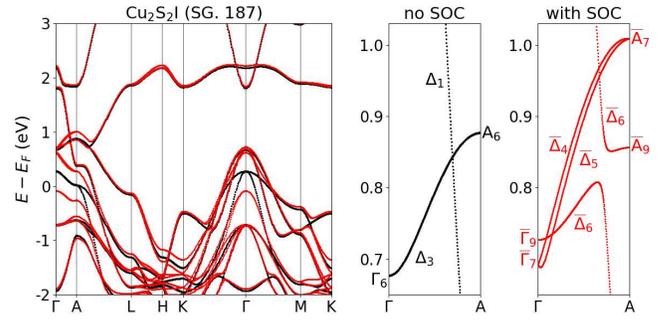} 
\caption{Electronic band structure of Cu$_2$S$_2$I without SOC (black) and with SOC (red) calculated. The bands are calculated with the PBE XC functional.}
\label{FigCu2S2I_bands}
\end{center}
\end{figure}

Cu$_2$S$_2$I crystallizes in the hexagonal SG 187. This symmetry group resembles SG 166 but it does not have inversion symmetry. We present the electronic band structure of Cu$_2$S$_2$I in Fig.~\ref{FigCu2S2I_bands}. As in the case of Cu$_2$Se$_2$I, there is a band crossing along $\mathbf{k}=(0,0,k_z)$ which is the $\Gamma$-A line in the BZ of SG 187. Without the inclusion of SOC, we observe that the twofold degenerate $\Delta_3$ band crosses the nondegenerate $\Delta_1$ band, similarly to the previous case of Cu$_2$Se$_2$I. When SOC is switched on, the $\Delta_1$ IRREP becomes $\overline{\Delta}_6$ (twofold degenerate). Interestingly, the $\Delta_3$ band splits into the twofold degenerate $\overline{\Delta}_6$ band and the nondegenerate $\overline{\Delta}_4$ and $\overline{\Delta}_5$ bands. The presence of nondegenerate bands in materials with time-reversal symmetry is possible only when inversion symmetry is broken, as is the case of SG 187. The $\Gamma$-A line in Cu$_2$S$_2$I thus features crossings of twofold degenerate bands with nondegenerate bands. Such crossings are called triple points~\cite{PhysRevX.6.031003} and their presence is protected by the symmetry of the material, similarly to the Dirac points.

While the $R-3m$ phase of CuI is a true topological semimetal, with Dirac points at the Fermi level, the symmetry-protected crossings in Cu$_2$Se$_2$I and Cu$_2$S$_2$I are located $\sim$1~eV above it. Moreover, there are also other electronic bands at the energy of the Dirac- and triple points in Cu$_2$Se$_2$I and Cu$_2$S$_2$I. These bands belong to a small hole pocket around the A point and thus we expect specific transport properties of the crossing points to be visible when the chemical potential is shifted to their energy.

These results were obtained with PBE. The authors of Ref.~\onlinecite{Le_2018} checked that for the $R-3m$ phase of CuI HSE03 band structures reduce the magnitude of the band inversion from 0.77~eV to $\sim$0.1~eV, but the topological Dirac semimetal character of the material remained preserved. We calculated the band structure of Cu$_2$Se$_2$I and Cu$_2$S$_2$I with HSE06 and HSE06 with modified $\alpha=0.32$ for comparison. In the case of Cu$_2$Se$_2$I, the band inversion and the Dirac points are preserved with HSE06, but HSE06 with modified $\alpha=0.32$ restores the band order and introduces a tiny band gap at T. For Cu$_2$S$_2$I we observe a small gap opening already with HSE06 and its size increases when the mixing $\alpha$ is set to 0.32. The comparison of the band structures obtained with different XC functionals is shown in Fig.~S5 of the SI.

In conclusion, we found two topological semimetals similar to the $R-3m$ phase of CuI which is reported to be a Dirac semimetal~\cite{Le_2018}. However, according to the materials project, the latter CuI phase is 256~meV/atom above the hull~\cite{Materialsproject_CuI_R-3m}, and therefore it will be quite a challenge to fabricate this structure. However, as we pointed out, the two phases we discussed here (Cu$_2$Se$_2$I and Cu$_2$S$_2$I) are on or very close to the hull. Thus, the incorporation of Se or S stabilizes the structure considerably. The new phases we predicted here could be a starting point for further investigations on topological CuI-based materials.

\subsection{Indirect semiconductors with chalcogen-chalcogen bonds}
To the third family of Cu-I-based ternaries includes indirect semiconductors with chalcogen-chalcogen bonds. In particular, we present here CuS$_2$I and CuSe$_2$I. Both crystals have an indirect band gap, and a notably smaller band gap than CuI (compare Table~\ref{Table_Ternaries_BandGaps}). They also possess effective masses larger than pristine CuI. We can conclude that these systems are not preferable to CuI as \textit{p}-type transparent conducting materials. The main difference compared to $\gamma$-CuI is the chalcogen-chalcogen bonds, which are of covalent nature. This counteracts the strong $p-d$ hybridisation present in zincblende-type structures, as presented in Table~\ref{Table_Ternaries_BandGaps}, which is an important ingredient for the p-type conductivity in CuI~\cite{Cardona_1963, Grundmann_2013, Williamson_2017}.

\subsection{Miscellaneous Structures}
The miscellaneous structures of the last family are unsuitable as a $p$-type TCM either due to lack of transparency or too low hole mobility. It might be interesting to describe in more detail Cu$_3$SeI$_2$, which possesses a layered structure. Compared to Cu$_3$SI$_2$,  the VBM is higher above the Fermi level and compensating doping might make harder to achieve. The band structures of these materials can be found in the SI. Cu$_3$SeI has a small effective mass but a too small band gap. On the other hand, CuSe$_3$I has a very large effective mass and a clear smaller admixture of $d$-states at the VBM. Our calculations support the importance of a strong $p-d$ hybridisation for the formation of $p$-type TCMs as discussed in the literature.

\section{Conclusion} 
Our accurate and extensive \textit{ab initio} calculations provide important insights into the ternary phase diagrams of Cu--S--I and Cu--Se--I. 
Our analysis reveals 11 ternary phases, 9 of which have not been reported before, that are on or very close to the convex hull of stability. We carefully characterized the electronic properties of these ternary compounds to identify potential $p$-type transparent conductive materials. Our results indicate that zincblende-type structures with the chalcogen atom partially replacing iodine on the anion sublattice show promise as the most suitable candidates, due to the possibility to easily control the carrier concentration through the Cu/I ratio, compared to pure $\gamma$-CuI. A strong $p-d$-hybridisation is present in those ternary compounds. Moreover, we identified Cu$_2$Se$_2$I and Cu$_2$S$_2$I as topological semimetals, with similar structural and electronic properties to a topological phase of CuI reported in the literature~\cite{Le_2018}. In contrast to the CuI topological phase, these ternary crystals are thermodynamically stable, providing an experimental basis for further investigations of topological CuI-based materials. Our results thus demonstrate that chalcogen doping and alloying can offer new strategies for improving and fine-tuning the electronic properties of copper iodide for real use in transparent electronics applications.

\begin{acknowledgments}
This work received funding from the Deutsche Forschungsgemeinschaft (DFG) through the research unit FOR~2857 and the projects BO~4280/9-1  and RA~3025/2-1. Computational resources were provided by the Leibniz Supercomputing Centre on SuperMUC (project pn68le).
\end{acknowledgments}

\section*{Data availability }
The data that support the findings of this study are available within this article and its supplementary material or are available from the corresponding authors upon reasonable request. Furthermore, the structures will be made available in future updates of Ref.~\cite{Schmidt_2022}.

\bibliography{CuITernaries}

\end{document}